\documentclass{ws-p8-50x6-00}
\def\R{{\cal R}}

\def\beq{\begin{equation}}
\def\eeq{\end{equation}}
\def\ba{\begin{eqnarray}}
\def\ea{\end{eqnarray}}

\def\yt{\widetilde y}

\def\Q2t{{\widetilde Q}^2}

\def\k0t{\widetilde{k}_0}

\def\magk{|{\bf k}|}
\def\magq{|{\bf q}|}
\def\lsim{\mathrel{\rlap{\lower4pt\hbox{\hskip1pt$\sim$}}
                   \raise1pt\hbox{$<$}}}
\def\gsim{\mathrel{\rlap{\lower4pt\hbox{\hskip1pt$\sim$}}
                   \raise1pt\hbox{$>>$}}}

\def\sf{spectral function}

\def\ci{\cite} 

\begin{document}

\title{A different view of deep inelastic electron-proton scattering}

\author{Omar Benhar}

\address{INFN, Sezione Roma 1. Piazzale Aldo Moro, 2. I-00185 Roma, Italy
\\E-mail: benhar@roma1.infn.it}


\maketitle

\abstracts{
Deep inelastic electron-proton scattering is analyzed in the
target rest frame using a theoretical approach suitable to describe many-body 
systems of {\em bound} constituents subject to {\em interactions}. 
At large three-momentum transfer $\magq$, this approach predicts the 
onset of scaling in the variable $\yt=\nu$-$\magq$, where $\nu$ denotes the 
energy transfer. The present analysis shows that the data, plotted at constant 
$\magq$, exhibit a remarkable scaling behavior in $\yt$ and manifestly 
display the presence of sizable interaction effects.
}

\section{Introduction}
The inclusive cross section for deep inelastic scattering (DIS) of 
unpolarized 
electrons by unpolarized protons can be written in terms of two structure
functions, $W_1$ and $W_2$, according to\cite{Halzen84}
\beq
\frac{d^2\sigma}{d\Omega d\nu} = \sigma_M \left[ W_2(|{\bf q}|,\nu) + 
2 W_1 (|{\bf q}|,\nu) \tan^2 \frac{\theta}{2}  \right]\ .
\label{eq:cs}
\eeq
In the above equation $\theta$ is the electron scattering angle, ${\bf q}$ and
$\nu$ denote the three-momentum and energy transfer, and
the Mott cross section is defined as 
$\sigma_M$~=~$\alpha^2\cos^2(\theta/2)/[4E_0^2 \sin^4(\theta/2)]$, $\alpha$ 
and $E_0$ being the fine structure constant and the beam energy, respectively. 

In general, $W_1$ and $W_2$ depend upon both 
${\bf q}$ and $\nu$. However, at large $Q^2$ ($Q^2=\magq^2-\nu^2$) 
 $F_1=mW_1$ and $F_2=\nu W_2$
have been observed to scale, i.e. to depend primarily on the Bjorken variable 
$x=Q^2/2(pq)$, where $p$ and $q$ denote the four momentum carried by the 
target proton and the four momentum transfer, respectively. In the target 
rest frame $x=Q^2/2m\nu$, m being the proton mass. 
The physical interpretation of the Lorentz scalar variable $x$ 
becomes apparent in the infinite momentum frame, defined by the condition 
${\bf p}\rightarrow\infty$, where it can be identified
with the momentum fraction carried by the proton constituent involved in the 
scattering
process\cite{Halzen84}. 

\begin{figure}[ht]
\begin{center}
\epsfxsize=18pc
\epsfbox{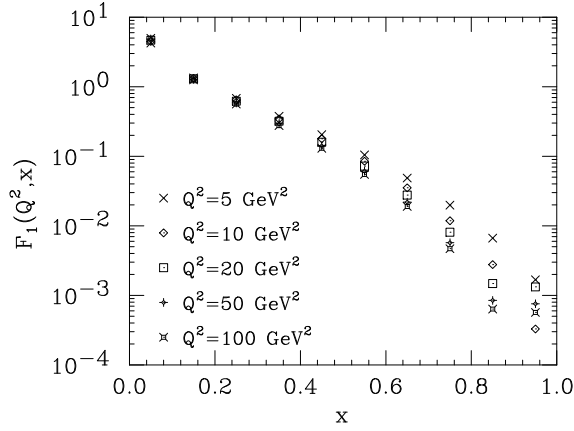}
\end{center}
\caption{
Proton $F_1$ obtained from the MRS(A) parton distributions
\protect\cite{Martin95}, plotted as a function of the Bjorken scaling 
variable $x$ for a range of values of $Q^2$
\label{norm:scal}}
\end{figure}

The scaling behavior of the observed proton $F_1$ is illustrated in 
fig. \ref{norm:scal}. The weak dependence upon $Q^2$, mostly due to 
gluonic radiative corrections, is well described by QCD  
evolution equations\cite{Ellis96}.  

In this note I discuss an approach to DIS based on many body theory, in which
proton constituents are treated as pointlike {\em bound} particles. 
Within this framework, the proton response is studied at constant $\magq$ 
as a function of the energy transfer $\nu$. This analysis naturally leads to 
predict the onset of scaling in the new variable $\yt = \nu - \magq$ at 
large $\magq$. 

Section 2 describes the many body treatment of the proton response at large
momentum transfer, the basic assumptions underlying the Plane Wave Impulse 
Approximation \mbox{(PWIA)}, and the emergence of the new variable $\yt$, 
closely related to the standard scaling variable of many body theory, $y$. 
In Section 3 the proton $W_1$ obtained from the MRS(A) parametrization of 
parton distributions\cite{Martin95} is shown to scale in $\yt$ and 
exhibit large effects driven by the binding energy of proton constituents.
Finally, section 4 is devoted to a summary and a discussion of the 
differences between the present approach and the standard parton model
description of DIS. The implications of the results presented in this
paper for the interpretation of different DIS observables are also briefly
analyzed. 

\section{Proton response within many-body theory and $\yt$-scaling}

The proton response to a scalar probe
delivering a {\em large} momentum ${\bf q}$ can be written
\beq
S({\bf q},\nu) = \sum_{F} | \langle F | \sum_i {\rm e}^{i{\bf q}\cdot{\bf r}_i}
 | 0 \rangle |^2\ \delta(\nu + m - E_F)\ ,
\label{s:scal}
\eeq
where ${\bf r}_i$ specifies the position of the i-th proton constituent. The 
above equation shows that $S({\bf q},\nu)$ is given by the
distribution of the strength of the state
$\sum_i i{\rm e}^{i{\bf q}\cdot{\bf r}_i}|0\rangle$,
created by the probe, among the eigenstates $|F\rangle$ of the system 
belonging to momentum ${\bf q}$.    

\begin{figure}[ht]
\begin{center}
\epsfxsize=18pc
\epsfbox{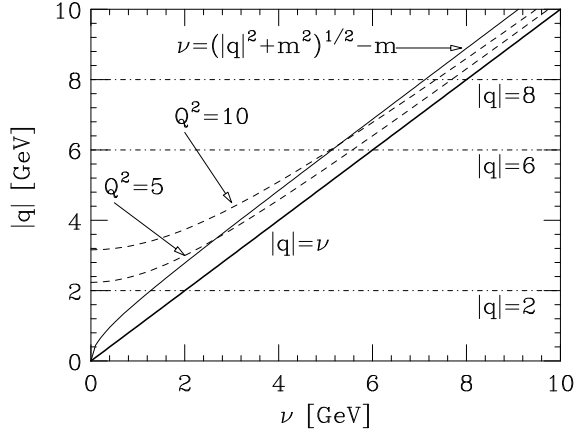}
\end{center}
\caption{
The thick solid line separates the spacetime and timelike
regions. The thin solid line corresponds to elastic electron-proton 
scattering, while the dashed lines correspond to constant $Q^2$. 
\label{qnuplane}}
\end{figure}

In general, the response (\ref{s:scal}) is nonzero at both $\nu \leq \magq$ and 
$\nu > \magq$. As a consequence, in many-body theory $S({\bf q},\nu)$ is 
usually studied at fixed $\magq$ as a function of the energy transfer $\nu$. 
The difference between this type of analysis and the one carried out at 
fixed $Q^2$ as a function of $x$ is illustrated in fig. \ref{qnuplane}. The 
dashed lines show the parabolae $Q^2$ = 5 and 10 GeV$^2$. They intersect the 
curve $\nu=\sqrt{\magq^2+m^2}-m$, 
 corresponding to elastic scattering kinematics, at $\nu=Q^2/2m$ (i.e. $x=1$) 
and approach the thick solid line $\nu=\magq$, separating the spacelike 
($\nu < \magq$) and timelike ($\nu > \magq$) regions as 
$\nu \rightarrow \infty$ (i.e. $x \rightarrow 0$). The standard analysis
of the structure functions is performed along these parabolae, 
that do not enter the timelike region. In the present paper I will discuss
the behavior of the proton response along the horizontal 
dash-dot lines of constant $\magq$, which extend into the $\nu > \magq$
region.

At large momentum transfer, scattering off a many-body system can 
be described as the incoherent sum of elementary processes 
in which the momentum tranfer ${\bf q}$ is delivered to the i-th constituent, 
carrying initial momentum ${\bf k}$, while the residual system, acting as 
a spectator, is left in the state $|\R\rangle$ with total momentum $-{\bf k}$. 
In PWIA, i.e. neglecting final state 
interactions (FSI) between the struck constituent and the residual system, 
the energy of the state 
\beq
|F\rangle = |i({\bf k}+{\bf q}); \R(-{\bf k})\rangle\ 
\label{final:state}
\eeq
can be written ($k_z={\bf k}\cdot{\bf q}/\magq$)
\beq
E_F = \magq + k_z + E_\R + O\left(\frac{1}{\magq}\right)\ .
\eeq
Combinig the requirement of energy conservation, $\nu+m=E_F$, and the above 
equation we obtain, in the $\magq \rightarrow \infty$ limit:
\beq
\yt = \nu - \magq = k_z + E_\R - m\ ,
\label{def:yt}
\eeq
implying in turn that as $\magq \rightarrow \infty$ 
$S({\bf q},\nu) \rightarrow S(\yt)$, i.e. that $S({\bf q},\nu)$ scales 
in the new variable $\yt$. 

The above discussion can be easily generalized to the case of electron scattering.
It should kept in mind, however, that $W_1$ and $W_2$ extracted from the 
measured electron-proton scattering cross section are not trivially 
related to the proton 
response. For example, while the proton response is in general nonvanishing in the 
$\nu \geq \magq$ region, at $\nu=\magq$ $W_2$ vanishes due to gauge invariance and
$W_1$ does not contribute to the cross section (see eq.(\ref{eq:cs})).          
In this note I will focus on $W_1$ only.

The expression of $W_1$ within PWIA can be easily obtained from 
eq.(\ref{s:scal}) replacing\cite{Benhar2000} 
\beq
\sum_{F} |F \rangle\langle F| \rightarrow \sum_{\R}\ \int\ d^3k\ 
|i({\bf k}+{\bf q}); \R(-{\bf k})\rangle
\langle\R(-{\bf k}); i({\bf k}+{\bf q})|\ ,
\eeq
and including the transverse cross section for electron scattering off a 
bound pointilke constituent carrying spin 1/2, defined as 
$\sigma_T=\sigma_M\sigma_1$. The results is
\beq
W_1(\magq,\nu)=\sum_{i}\ \int\ d^3k\ dE\ \sigma_1({\bf k},E,{\bf q},\nu)\ 
P_i({\bf k},E)\ \delta(\yt-k_z-E)\ ,
\eeq
where 
\beq
P_i({\bf k},E) = \sum_{\R}\ |\langle \R(-{\bf k}); i({\bf k})|0\rangle|^2\ 
\delta(m-E_\R-E)\ .
\eeq
In the limit of vanishing constituent mass $\sigma_1 \rightarrow e_i^2$, 
$e_i$ being the electric charge of the i-th constituent, and we get
\beq
W_1(\magq,\nu) = 
\sum_{i}\ e_i^2\int\ d^3k\ dE\ P_i({\bf k},E)\ \delta(\yt-k_z-E)\ =
\sum_{i}\ e_i^2\ {\widetilde f}(\yt)\ ,
\eeq
showing that i) the structure function $W_1$ scales in $\yt$ and ii) it 
provides 
a \mbox{direct} measurement of the response. It is apparent that $\yt$ closely
resembles the scaling variable $y$ used in  quasi-elastic electron-nucleus
scattering, where it is associated with the component of the momentum of the
struck nucleon\cite{Day90} parallel to the momentum transfer. However, it 
should be pointed out that since 
$\yt = -m \xi$, where $\xi$ is the Nachtmann \cite{Nachtmann73,Jaffe85}
 variable, $\yt$-scaling is related to Bjorken scaling as well.    

The simple picture of the proton response described so far will be 
obviously modified by 
color confining interactions. While the mass of the nucleon contains the 
contribution of confinement interactions, this
contribution is omitted in the energy of the 
struck quark, $\magq + k_z$. Therefore, it must be included in the energy $E(\R)$
of the residual system. We expect that the confinement energy does not
change significantly in the time duration of DIS, and its main
influence is via the wave functions $|0\rangle$ and $|\R\rangle$.
However, it could also contribute to FSI.

\section{Data analysis}

\begin{figure}[ht]
\begin{center}
\epsfxsize=18pc
\epsfbox{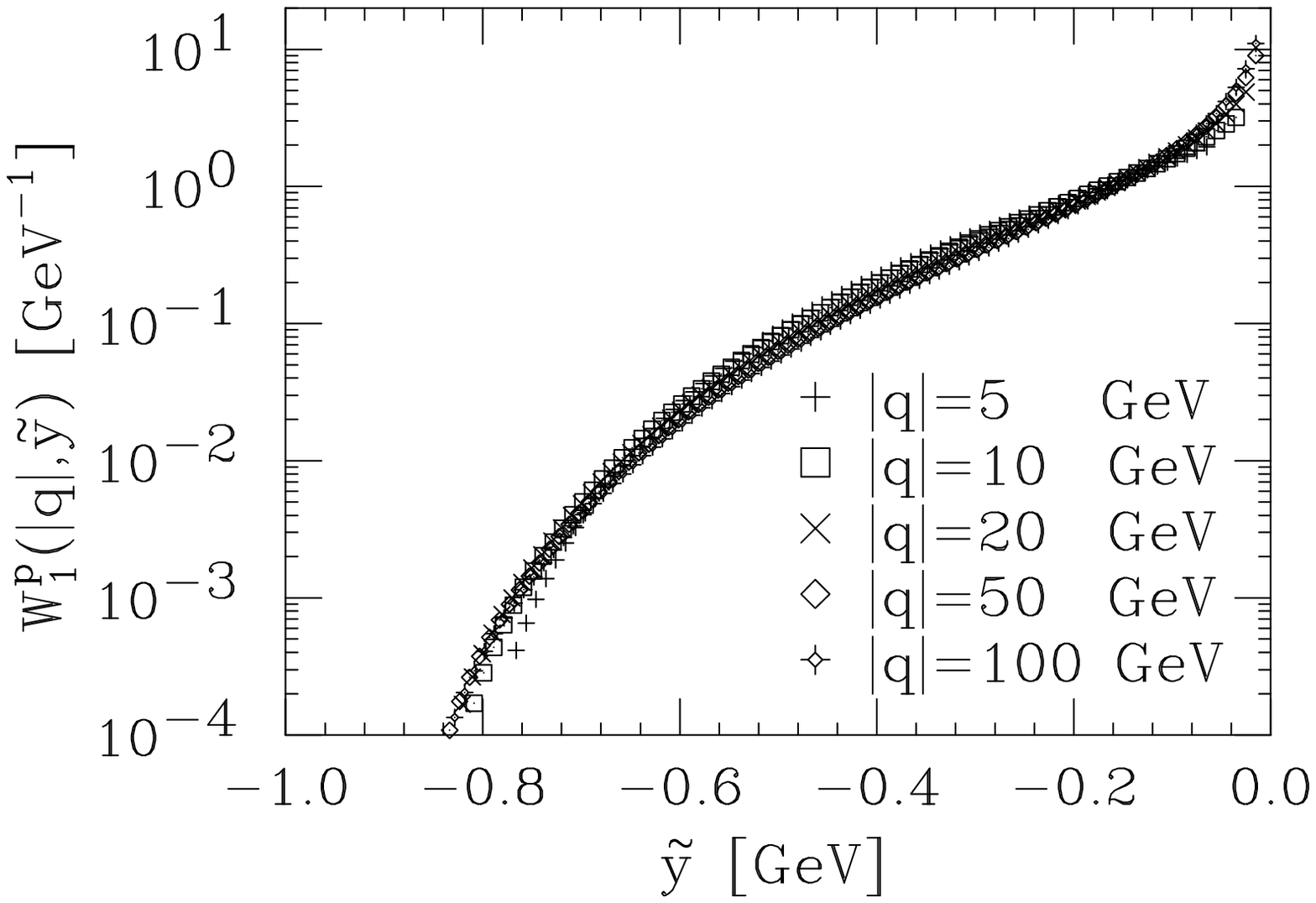}
\end{center}
\caption{
Proton W$_1$, obtained from the MRS(A) parton distributions, plotted as a function 
of the scaling variable $\yt = \nu - \magq$ for a range of values of $\magq$.
\label{w1pqnu}}
\end{figure}

Fig. \ref{w1pqnu} shows the proton $W_1({\bf q},\nu )$, obtained from
the MRS(A) fit of parton distributions\cite{Martin95}, plotted at several
values of $|{\bf q}|$ as a function of $\yt$.
The MRS(A) parton distributions provide a smooth description of the data
on the structure function $F_2(Q^2,x)$ all the way from $Q^2=10^{-1}$ GeV
to $Q^2=10^{3}$ GeV. It clearly appears that at large values of
 ${\magq}$ $W_1({\bf q},\nu )$ depends primarily on $\yt$.
As pointed out in the previous section, this scaling has a simple interpretation 
within many-body theory, related to the well known $y$-scaling\cite{Day90}
observed in electron-nucleus scattering.

The small scaling violations seen in fig. \ref{w1pqnu}  can be ascribed to 
gluonic radiative corrections, as in the standard $x$-scaling analysis.
Note however that in $\yt$-scaling  $Q^2$ has a large variation, 
ranging from zero at $\magq = \nu $ ($\yt=0)$ to $\sim$ 2m$\magq$ at 
$\nu=\magq-m$ ($\yt=m$), which does not appear to spoil the quality 
of scaling.

An interesting feature of fig. \ref{w1pqnu} concerns the width
of the response, which turns out to be few hundred MeV only, independent 
of the value of $\magq$. This implies that DIS
 has an intrinsic energy scale of few hundred MeV.
The main part of the energy tranfer $\nu$, of order $\magq$, goes into the
kinetic energy of the struck constituent, and does not play any interesting
role in the dynamics of the target system. It follows that changes in the
energy $E_\R$ of the residual system of order 100 MeV 
significantly affect $W_1(\yt)$. 

The effects of $E_R$ on the response can be clearly seen in the following 
example, first sudied by Close and Thomas\cite{Close88}.
Let us consider the difference
between the  responses due to valence u and d quarks in the proton. 
 When the
electron strikes the valence d quark, the remaining two valence u quarks are
left in the residual state $\R_1$ with spin 1. On the other hand, when a 
valence u quark is struck, the residual ud pair can be found in the 
spin 0 state, 
$\R_0$, with probability 0.75, or in the spin 1 state, $\R_1$, with probability 
0.25. Denoting by $V_u(\yt )$ and $V_d(\yt )$ the contributions of
valence u and d quarks to $W_1(\yt)$ we can write the response due to the final 
state $\R_1$ (normalized to unit constituent charge)
\beq
\chi_1(\yt) = 9 V_d(\yt)\ ,
\label{chi:0}
\eeq
while that associated with $\R_0$ is
\beq
\chi_0(\yt) = \frac{3}{2} \left[ V_u(\yt)-2 V_d(\yt) \right]\ .
\label{chi:1}
\eeq
In perturbation theory 
\beq
E_{{\R}_1} - E_{{\R}_0} \sim \frac{2}{3}(m_{\Delta}-m) \sim 0.2\ {\rm GeV}\ ,
\eeq
and $\chi_1(\yt)$ is expected to be shifted to larger $\yt$ by $\sim 0.2$ GeV, 
 with respect to $\chi_0(\yt)$.

\begin{figure}[ht]
\begin{center}
\epsfxsize=18pc
\epsfbox{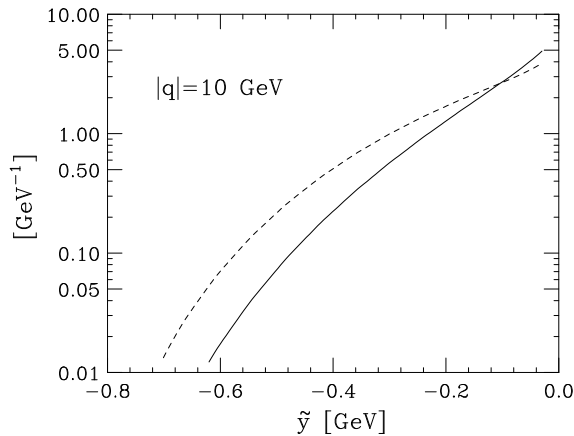}
\end{center}
\caption{
Proton valence quark responses $\chi_0$ (dashed) and $\chi_1$ (solid),
defined in eqs.(\protect{\ref{chi:0}}) and (\protect{\ref{chi:1}}), at
 $\magq=10$ GeV. 
\label{ud}}
\end{figure}

 Fig. \ref{ud} shows that $\chi_0$ and $\chi_1$ obtained from the MRS(A) parton 
distributions at
$\magq = 10$ GeV are indeed shifted by $\sim 0.1$ GeV from each other
at $\yt < -0.2$ GeV. According to PWIA this shift should 
be independent of $\yt$, 
provided the color magnetic interaction can be treated perturbatively. 
The fact that the shift is only $\sim$ 0.1 GeV indicates that it has 
nonperturbative contributions. Differences in FSI can also have an influence.          
\section{Summary and conclusions}

The results presented in this paper show that the structure function 
$W_1$, which is proportional to the proton response in the limit of 
vanishingly small constituent mass, scales in the impulse approximation 
variable $\yt$ at large $\magq$. The occurrence of $\yt$ scaling emerges 
in a most natural way from the many body treatment of DIS in the target 
rest frame. 

Unlike the standard parton model of DIS, many body theory 
 treats proton constituents as {\it bound} particles. This difference 
has a number of relevant implications. 
As pointed out in Section 2, within the present approach the response is
in general nonzero in the $\nu > \magq$ region, not accessible by electron 
scattering. A timelike response can occur either due to initial
state interactions, which  can make $E_\R$
large enough to give a positive right hand side of eq.(\ref{def:yt}),
 or because of FSI. The initial energy of the struck
constituent is identified with $e = m - E_\R$
 and not with the on-shell energy $\sqrt{m_i^2 + \magk^2}$, where 
$m_i$ is the mass of the i-th constituent. 
On the other hand, in the parton model 
the struck particle is assumed to be on mass-shell before and
after the interaction with the electron \cite{Ellis83}. In this case
\beq
\nu = \sqrt{m_i^2+({\bf k}+{\bf q})^2}-\sqrt{m_i^2+{\bf k}^2} \leq \magq\ ,
\label{eq:nuparton}
\eeq 
and all of the response is at negative $\yt$, in the spacelike region. 
The same conclusion applies to the leading twist-two order of the operator 
product expansion \ci{Jaffe85}.    
The fact that putting the quarks on mass-shell, as commonly done
in perturbative QCD, is {\it not legal} has been recently pointed out 
by Bjorken\cite{Bjorken2000}. 


In conclusion, new insights in DIS off the proton can be obtained
using standard many-body theory and relating the scaling
function to the distribution of proton constituents in the 
proton rest frame. While $\yt$ scaling is derived assuming {\it bound}
constituents subject to {\it interactions}, the occurrence of 
$\xi$- or $x$-scaling is obtained under the assumption of free, on-shell, 
constituents. Thus, the occurrence of scaling cannot automatically be taken 
as evidence of scattering off free constituents.        

\section*{Acknowledgments}
The work described in this paper has been done in collaboration 
with V.R.~Pandharipande and I. Sick.

\end{document}